\documentclass{aa}
 
\begin{document}
 
\thesaurus{07 
                  (07.13.1; 
                   07.13.2)} 
 
\title{A solution for the Tunguska event}
 
\author{Luigi Foschini}
 
 
\institute{CNR - Institute FISBAT, Via Gobetti 101, I-40129 Bologna,
Italy; (email: L.Foschini@fisbat.bo.cnr.it)}
 
\date{Received 1 September 1998; accepted 10 December 1998}
 
\titlerunning{A solution for the Tunguska event} \authorrunning{L.
Foschini}
 
\maketitle
 
\begin{abstract}
This letter presents a new solution for the Tunguska event of June
30th, 1908.  The solution has been obtained starting from seismic
data, is in fair agreement with the observational evidence, and
supports the asteroidal hypothesis for the origin of the Tunguska
cosmic body.  It is based on an improved model of the hypersonic flow
around a small asteroid in the Earth's atmosphere.
\end{abstract}
 
\section{Introduction}
On June 30th, 1908, something exploded over Tunguska, in central
Siberia.  Over the last ninety years this catastrophic event has
inspired a plethora of scientific investigations. Despite many
interesting findings, there are still substantial open
questions and inconsistencies among the theories and the available
data (for a review see Vasilyev \cite{VASILYEV}).
 
Among many different effects, the Tunguska explosion produced
shock waves, which were recorded by seismographs
at several sites.  Ben--Menahem (\cite{MENAHEM}) made a detailed
analysis of these seismic records and derived an explosive energy of
$12.5\pm 2.5$ Mton.  He also concluded that the data on the energy
source are consistent with an airburst at a height of about 8.5~km.
 
In a previous letter (\cite{FOSCHINI}), we have shown that seismic
data can be used to characterize the very bright 1993 Lugo bolide,
obtaining a good match between the derived solution and the
observations.  Here the same methodology is applied to analyze the
Tunguska event, using Ben--Menahem's analysis as a starting point.
 
\section{Current models}
Several different models have been developed in order to fit all the
available data on the Tunguska event (e.g.  Chyba et al. \cite{CHYBA},
Grigorian \cite{GRIGORIAN}, Hills \& Goda \cite{HILLS}, Lyne et al.
\cite{LYNE}).  All these models have contributed significantly to a
general improvement in our understanding of the atmospheric disruption
of meteoroids.  They usually assume that the fragmentation process
starts when the aerodynamic pressure is equal to the mechanical
strength $S$ of the cosmic body.  Relating air density to airburst
height, this allows one to derive the meteoroid speed ($V$):

\begin{equation}
   V=\sqrt{\frac{S}{\rho_{\mathrm{sl}}}\exp\left[\frac{h}{H}\right]}\, ,
   \label{e:velo}
\end{equation}

\noindent where $\rho_{\mathrm{sl}}$ is the atmospheric density at sea level,
$h$ is the height of first fragmentation and $H$ is
the atmospheric scale height (about 8~km).  From Ben--Menahem's
analysis we infer that there was a single fragmentation event; there
is no evidence of multiple explosions, as it should occur during
multiple fragmentation events (Ben--Menahem \cite{MENAHEM}).  Thus
Eq.~(\ref{e:velo}) can be used to derive $V$, provided one assumes
that the first fragmentation coincided with the airburst occurred at $h
= 8.5$~km.
 
For different types of cosmic body, corresponding to different assumed
values for $S$ (taken from Hills \& Goda, \cite{HILLS}), we obtain the
results listed in Table~\ref{speed-actual}.
 
\begin{table}[h]
    \centering
    \caption{Speed of the Tunguska cosmic body vs. strength according
    to Eq.~(1)}
    \begin{tabular}{lcc}
    \hline
    Body type & $S$ [Pa] & $V$ [km/s]\\
    \hline
    Comet & $1\cdot 10^{6}$ & $1.5$\\
    Carbonaceous chondrite & $1\cdot 10^{7}$ & $4.7$\\
    Stone & $5\cdot 10^{7}$ & $10.6$\\
    Iron & $2\cdot 10^{8}$ & $21.2$\\
    \hline
    \end{tabular}
    \label{speed-actual}
\end{table}
 
\noindent Now, since before exploding large meteoroids undergo a
limited mass loss during their atmospheric path, the pre--explosion
speed must be close to the (geocentric) orbital speed, and thus must
be greater than the Earth's escape velocity ($11.2$~km/s).  Therefore,
according to the results derived from Eq.~(1), the most plausible
solution would be that of an iron body.  However, the iron body
hypothesis is not consistent with the recent on--site recovery of
microremnants from a stony object (Longo et al. \cite{LONGO}, Serra et
al. \cite{SERRA}).
 
Actually, taking into account the uncertainty in the value of $S$ and
the different measurement errors (both of which are difficult to
quantify), the stony object solution could not be entirely ruled out
using this argument (typical geocentric speeds for near--Earth
asteroids are $\approx 15$~km/s).  However, it is known that the
interaction of large meteoroids (or small asteroids) with the Earth's
atmosphere is characterized by a great variety of behaviors, and any
quantitative theory should take into account a large number of
variables: size, shape, rotation, composition, internal structure,
orbital speed, flight path angle.  Thus for the time being each bolide
must be seen as a case study, which can provide useful insights for a
future comprehensive theory.  As a further consequence, Eq.~(1)
cannot be trusted to provide quantitatively reliable results in every
case.
 
For instance, we know that sometimes meteoroids explode at dynamic
pressures much lower than their mechanical strength (Ceplecha
\cite{CEPLECHA2}).  In the case of the Lugo bolide, an interesting
possibility is that this behavior may have been related to a porous
structure of the meteoroid (Foschini \cite{FOSCHINI}).  However,
Tab.~\ref{speed-actual} shows that in the case of Tunguska we have the
opposite problem, and that we should assume an anomalously high
mechanical strength.  Therefore, I will look into another direction
for a possible solution of the conundrum.
 
\section{Hypersonic flow}
When a large meteoroid enters the Earth's atmosphere, it has a speed
in the range $12\div 72$~km/s, hence it moves at \emph{hypersonic}
speed (that is, with Mach number greater than about 5).  Since here we
are interested in the dynamics of a meteoroid large enough to reach
the lower atmosphere, the fluid can be treated as a continuum.  Thus,
we can use current knowledge about hypersonic aerodynamics in order to
understand meteoroid airbursts.  For thorough presentations of this
theory, the reader is referred to the books of Shapiro
(\cite{SHAPIRO}), Landau \& Lifshitz (\cite{LANDAU}), and Holman
(\cite{HOLMAN}).
 
It is important to note that for large Mach numbers the linearized
equations for the speed potential are not valid, so we cannot use laws
holding for supersonic speeds.  In hypersonic flow, Mach waves and
oblique shock waves are emitted at small angles with the direction of
the flow, of the order of the ratio between body thickness and length,
and thus tend to follow the surface of the body.  Under these
conditions, the atmospheric path of a large meteoroid can be seen as a
long cylinder, generating pressure waves that can detected as
infrasonic sound (Cumming \cite{CUMMING}, ReVelle \cite{REVELLE}).
 
The small angle of Mach and oblique shock waves gives also rise to the
concept of hypersonic boundary layer near the surface.  In front of
the meteoroid there is a bow shock, that envelopes the body.  The
shock is stronger on the symmetry axis, because at this point it is
normal to the stream.  Then, we find a zone where molecular
dissociation is the main process and even closer to the body surface,
we find the boundary layer, where viscous effects are dominant.  As
the air flows toward the rear of the meteoroid, it is reattracted to
the axis, just like in a Prandtl--Meyer expansion.  As a consequence,
there is a rotation of the stream in the sense opposite to that of the
motion (rectification); this creates an oblique shock wave, which is
called wake shock.  Since the pressure rise across the bow shock is
huge when compared to the pressure decrease across the Prandtl--Meyer
expansion, one can assume, as a reasonable approximation, that there
is a vacuum in the rear of the meteoroid.  For illustrative images of
a hypersonic flow, we refer to Chapter 19, Volume 2, of Shapiro
(\cite{SHAPIRO}).
 
The fluid temperature increases in the boundary layer, because the
speed must decrease to zero at the meteoroid surface; moreover there
are heating effects due to viscous dissipation.  There are also
regions (like in the Prandtl--Meyer expansion) in which the presence
of vacuum or near--vacuum strongly reduces heat transfer, and this
contributes to the increasing body temperature.  If the generation of
heat increases so quickly that the loss of heat may be inadequate to
achieve an equilibrium state, we may have a thermal explosion.  This
explosion generates pressure waves that can be detected on the ground
by seismographs. Note that after the Tunguska event no meteorite was
recovered, so the argument the meteorites are usually cold immediately
after landing does not rule out this kind of thermal explosion in this
case.
 
Current models of the Tunguska event consider, as a reference, the
stagnation pressure only (e.g.  Hills \& Goda \cite{HILLS}), although,
for the reasons outlined above, a realistic physical description
should account for heat transfer and generation processes as well.  A
similar conclusion on the need for a coupled radiation--hydrodynamical
model has been recently reached by Borovi\v{c}ka et al.
(\cite{BORO1}, \cite{BORO2}), following a detailed analysis of
theories and observations for the Bene\v{s}ov bolide.
 
\section{The importance of the stagnation temperature}
Let us now consider the heating due to the conversion of kinetic
energy of the flow into thermal energy, when the gas is brought to
rest (in the boundary layer).  This process can be described in terms
of a steady flow energy in an adiabatic process:

\begin{equation}
h_{0}-h_{\infty}=\frac{V_{\infty}^{2}}{2}\, ,
\label{e:enthalpy}
\end{equation}

\noindent where $h_{0}$ and $h_{\infty}$ are the stagnation and free
stream enthalpy of the fluid, respectively, and $V_{\infty}$ is the
free stream speed.  Note that the choice of the reference frame is not
important here: if we consider a frame centered on the body, the fluid
will move, and {\it vice versa}; therefore $V_{\infty}$ can be
interpreted as the body's speed with respect to the atmosphere.
We can rewrite Eq.~(\ref{e:enthalpy}) in terms of temperature:

\begin{equation}
T_{0}-T_{\infty}=\frac{V_{\infty}^{2}}{2c_{\mathrm{p}}}
\label{e:temp}
\end{equation}

\noindent where $c_{\mathrm{p}}$ is the specific heat at constant pressure.
During the atmospheric path, as the Mach number is large, the
meteoroid's speed is close to the maximum value corresponding to the
stagnation temperature.  Changes in the stream properties are mainly
due to changes in the stagnation temperature $T_{0}$, which is a direct
measure of the amount of heat transfer.

This argument stresses the importance of the stagnation temperature in 
hypersonic flow, since it is related to the maximum speed of the stream, 
which in turn is close to the speed of the cosmic body. According
to Shapiro (\cite{SHAPIRO}), the relationship between stagnation
temperature and maximum speed of the stream can be expressed in the
following way:

\begin{equation}
V_{\mathrm{max}}=\sqrt{\frac{2\gamma}{\gamma-1}RT_{0}}\, ,
\label{e:vmax} 
\end{equation}

\noindent where $\gamma$ is the ratio of specific heats.  By means of the 
equation of state for the air, $V_{\mathrm{max}}$ can be expressed as a 
function of the stagnation pressure and density:

\begin{equation}
V_{\mathrm{max}}=\sqrt{\frac{2\gamma}{\gamma -1}\frac{p_{0}}{\rho_{0}}}\, .
\label{e:vmax2} 
\end{equation}

In order to obtain a condition for the
meteoroid breakup, the stagnation pressure $p_0$ must be set equal to
the mechanical strength $S$ of the body.  As for  the stagnation
density, we have $(\rho_{0}-\rho_{\mathrm{air}})/\rho_{\mathrm{air}}\approx 1$ (Landau
\& Lifshitz \cite{LANDAU}), where $\rho_{\mathrm{air}}$ is the undisturbed air
density at the airburst height.  Finally, by expressing $\rho_{\mathrm{air}}$
as a function of atmospheric height $h$ and $\rho_{\mathrm{sl}}$, like in
Eq.~(1), we obtain a new equation to estimate $V_{\mathrm{max}}$, which is
close to the speed of the cosmic body at breakup $V$:

\begin{equation}
V\approx V_{\mathrm{max}}=\sqrt{\frac{\gamma}{\gamma
-1}\frac{S}{\rho_{\mathrm{sl}}} \exp\left[\frac{h}{H}\right]}\, .
\label{e:velo2} 
\end{equation}

For $\gamma$ we can use a value of
about $1.7$, resulting from experimental studies on plasma developed
in hypervelocity impacts (Kadono \& Fujiwara \cite{KADONO}).
Comparing Eq.~(\ref{e:velo2}) to Eq.~(\ref{e:velo}), we see an
additional factor of about $1.6$.  This comes from the fact that
Eq.~(\ref{e:velo2}) derives from Eq.~(\ref{e:vmax}), according to
which the stagnation temperature depends on speed when a body is
travelling at hypersonic velocity.  Eq.~(\ref{e:velo2}) shows that the
airburst occurs thanks to the combined thermal and mechanical effects
acting on the meteoroid. In other words, thermodynamic processes
decrease the effective pressure crushing the body in a
significant way, so the same body can reach a lower altitude, or for
a given airburst altitude a lower strength is required.
 
\section{A new analysis of the Tunguska event}
By means of Eq.~(\ref{e:velo2}) we can replace Table ~1 with a new
table for the breakup speeds of different types of cosmic body (see
Tab.~\ref{speed-new}).  Note that now the inferred speed for an iron
body would be too high, and stony bodies provide the most plausible
solution.  This is consistent with the results of a detailed analysis
of several hundreds meteors carried out by Ceplecha \& McCrosky
(\cite{JGR}) and Ceplecha (\cite{CEPLECHA}), who found that a height
around 10~km is fairly typical for stony objects.
 
\begin{table}[h]
\centering
\caption{Speed of the Tunguska cosmic body vs. strength according
to Eq.~(\ref{e:velo2})}
\begin{tabular}{lcc}
\hline
Body Type & $S$ [Pa] & $V$ [km/s]\\
\hline
Comet & $1\cdot 10^{6}$ & $2.3$\\
Carbonaceous Chondrite & $1\cdot 10^{7}$ & $7.4$\\
Stone & $5\cdot 10^{7}$ & $16.5$\\
Iron & $2\cdot 10^{8}$ & $33.0$\\
\hline
\end{tabular}
\label{speed-new}
\end{table}

We can now calculate other data for the Tunguska event solving the
equations of motion and the luminosity equation, according to the
procedure described in Foschini (\cite{FOSCHINI}).  The results are
summarized in Table~\ref{summary}.  The following assumptions have
been made: (i) the luminous efficiency $\tau$ is $5\%$; (ii) the
diameter of the object is calculated assuming a spherical shape and a
density of $3500$~kg/m$^{3}$, typical for a stony object.

\begin{table}[h]
\centering
\caption{Summary on the properties of the Tunguska Cosmic Body}
\begin{tabular}{lr}
\hline
Apparition time (UT)$^{\mathrm{a}}$ & 1908 06 30 00:14:28\\
Latitude of airburst$^{\mathrm{a}}$ & $60\degr 55'$~N\\
Longitude of airburst$^{\mathrm{a}}$ & $101\degr 57'$~E\\
Airburst height$^{\mathrm{a}}$ & $8.5$~km\\
Explosion Energy$^{\mathrm{a}}$ & $12.5$~Mton\\
Mass & $4\cdot 10^{8}$~kg\\
Diameter & $60$~m\\
Abs.  Visual Magnitude & $-29.4$\\
Velocity & $16.5$~km/s\\
Inclination$^{\mathrm{b}}$ & $3\degr$\\
Path azimuth$^{\mathrm{a,c}}$ & $115\degr$\\
\hline
\end{tabular}
\label{summary}
\begin{list}{}{}
\item[$^{\mathrm{a}}$] From Ben--Menahem (\cite{MENAHEM}).
\item[$^{\mathrm{b}}$] Over the horizon.
\item[$^{\mathrm{c}}$] Clockwise from North.
\end{list}
\end{table}
 
Comparing these results to previous ones and to the available data
(for a review see Vasilyev \cite{VASILYEV}), we note a generally good
agreement, except for the trajectory inclination over the horizon.
The value obtained here is about $3\degr$, while Vasilyev
reported that the most likely inclination angle was
about $15\degr$.  However, he also noted the possibility of a good
aerodynamic shape of the Tunguska cosmic body, that may have decreased
the inclination angle.  Moreover, we have neglected the lift effects,
following Chyba et al. (\cite{CHYBA}).
 
Among the authors quoted by Vasilyev, only Sekanina derived an
angle lower than $5\degr$.  Interestingly, it was just Sekanina
(\cite{SEKANINA}) who strongly favoured the conclusion of an
asteroidal origin for the Tunguska cosmic body. The results obtained
here provide additional support for Sekanina's conclusion.
 
\section{Conclusions}
In this paper, we have outlined a new analysis of the the Tunguska
event, starting from seismic data obtained by Ben--Menahem
(\cite{MENAHEM}) and improving the relationship between body speed,
mechanical strength and airburst height.  The main conclusion is that
the Tunguska cosmic body was probably a stony asteroid, with a
diameter of about 60~m.
 
We have also summarized the properties of the hypersonic flow around a
small asteroid in the Earth's atmosphere.  We have shown that the
stagnation temperature is a direct measure of the body speed.  This
introduces a multiplicative factor of $\sqrt{\gamma/(\gamma-1)}$ in
the Eq.~(\ref{e:velo}), which is instrumental to derive a reasonable
solution for the Tunguska event.  Eq.~(\ref{e:velo2}) is consistent
with the idea that the meteoroid's fragmentation is due to the coupled
action of thermodynamical and mechanical processes.
 
This kind of analysis can be applied whenever the body is large and
compact enough to reach the lower atmosphere.  Further investigations
are needed for application to other cases.  However, the crucial
role of the stagnation temperature is probably a general
feature of any realistic model of meteoroid flight and breakup.
 
\begin{acknowledgements}
The author is grateful to an anonymous referee for useful comments. A special thank to
P.~Farinella for constructive review.
\end{acknowledgements}

\end{document}